\documentclass[preprintnumbers,amsmath,amssymb,mathabx,textcomp,showpacs,superscriptaddress,lengthcheck,floatfix,pre]{revtex4}

\usepackage{graphicx}% Include figure files
\usepackage{dcolumn}% Align table columns on decimal point
\usepackage{bm}% bold math

\usepackage{latexsym}
\usepackage{amsmath}
\usepackage{amssymb}
\usepackage{graphicx}
\usepackage{dcolumn}% Align table columns on decimal point
\usepackage{booktabs}
\usepackage{bm}
\usepackage{epsfig}
\usepackage{color}
\usepackage{graphicx}% Include figure files
\usepackage{multirow}
\usepackage{float}

\newcommand{\point}[1]{({#1})}

\newcommand{\KA}{*}
\newcommand{\cp}{{\textrm{cp}}}
\newcommand{\maxx}{{\textrm{max}}}

\newcommand{\Nparticles}{N}
\newcommand{\Ncp}{N_\textrm{cp}}
\newcommand{\junctions}{J}
\newcommand{\Klobe}{K}
\newcommand{\Mvoids}{M}
\newcommand{\MvoidsMax}{M_\maxx}

\newcommand{\km}{k_m}
\newcommand{\ke}{k_e}
\newcommand{\Nc}{{N^c}}
\newcommand{\Nci}[1]{N^c_{#1}}

\newcommand{\lobeLength}{L}
\newcommand{\channelLength}{L}
\newcommand{\intersectionLength}{l}
\newcommand{\particleLength}{d}
\newcommand{\lobeLengthE}{\lobeLength_e}
\newcommand{\lobeLengthM}{\lobeLength_m}
\newcommand{\lobeLengthMKA}{\lobeLengthM^\KA}

\newcommand{\phiKA}{\phi^\KA}
\newcommand{\phicp}{\phi^\cp}

\newcommand{\crossingExponent}{\alpha}

\newcommand{\squeezedSet}{\mathcal{Q}}
\newcommand{\unsqueezedSet}{\mathcal{U}}
\newcommand{\network}{\mathcal{N}}

\newcommand{\tauD}{\tau_0}

\newcommand{\occupancyVar}{\textbf{\textit{S}}}

\begin{document}

\title{Quasi-One Dimensional Models for Glassy Dynamics}
\author{Prasanta Pal}
\affiliation{ Department of Diagnostic Radiology, Yale University School of Medicine, New Haven, CT, 06520-8042 }
\author{Jerzy Blawzdziewicz}
\affiliation{Department of Mechanical Engineering, 
Texas Tech University, Lubbock, TX 79409-1021}
\author{Corey S. O'Hern}
\affiliation{Department of Mechanical Engineering \& Materials Science, Yale University, New Haven, CT 06520-8286}
\affiliation{Department of Applied Physics, Yale University, New Haven, CT  06520-8267}
\affiliation{Department of Physics, Yale University, New Haven, CT  06520-8120}

\begin{abstract}
We describe numerical simulations and analyses of a quasi-one-dimensional (Q1D)
model of glassy dynamics.  In this model, hard rods undergo Brownian
dynamics through a series of narrow channels connected by $J$
intersections. We do not allow the rods to turn at the intersections,
and thus there is a single, continuous route through the system.  This
Q1D model displays caging behavior, collective particle
rearrangements, and rapid growth of the structural relaxation time,
which are also found in supercooled liquids and glasses.  The
mean-square displacement $\Sigma(t)$ for this Q1D model displays several
dynamical regimes: 1) short-time diffusion $\Sigma(t) \sim t$, 2) a
plateau in the mean-square displacement caused by caging behavior, 3)
single-file diffusion characterized by anomalous scaling $\Sigma(t)
\sim t^{0.5}$ at intermediate times, and 4) a crossover to long-time
diffusion $\Sigma(t) \sim t$ for times $t$ that grow with the
complexity of the circuit.  We develop a general procedure for
determining the structural relaxation time $t_D$, beyond which the
system undergoes long-time diffusion, as a function of the packing
fraction $\phi$ and system topology.  This procedure involves several
steps: 1) define a set of distinct microstates in configuration space
of the system, 2) construct a directed network of microstates and
transitions between them, 3) identify minimal, closed loops in the
network that give rise to structural relaxation, 4) determine the
frequencies of `bottleneck' microstates that control the slow dynamics
and time required to transition out of them, and 5) use the microstate
frequencies and lifetimes to deduce $t_D(\phi)$.  We find that $t_D$
obeys power-law scaling, $t_D\sim (\phi^* -
\phi)^{-\alpha}$, where both $\phi^*$ (signaling complete kinetic arrest) and $\alpha>0$ depend on the system
topology.  

\end{abstract}
\date{\today}

\maketitle

\section{Introduction}
\label{intro}

Developing a fundamental understanding of glass transitions in
amorphous materials is one of the remaining grand challenges in
condensed matter
physics~\cite{sid,phasetransitiondynamics,chaikinbook}.  Glass
transitions occur in myriad systems including atomic, magnetic,
polymer, and colloidal systems.  Hallmarks of the glass
transition include a stupendous increase in the structural and stress
relaxation times~\cite{angell2} and a concomitant dramatic decrease in
the mobility over an extremely narrow range of temperature or density,
broad distributions of particle motions that are spatially and
temporally heterogeneous, and aging behavior in which the system
becomes progressively more viscous with time after it has been
quenched to the glassy state~\cite{debenedetti}.

Glass transitions in liquids show marked similarities to jamming
transitions in athermal systems such as granular media, foams, and
emulsions that do not thermally fluctuate~\cite{jam}.  Athermal
systems typically jam, or develop a nonzero static shear modulus, at
sufficiently large densities or confining pressures, and remain jammed
for applied shear stresses below the yield stress.  Similarities
between jammed and glassy systems include highly cooperative and
heterogeneous particle motion in response to
perturbations~\cite{dauchot,dauchot2} and extremely slow
relaxation~\cite{dauchot3} as a system approaches the glass or jamming
transition.

\begin{figure}
\begin{center}
\scalebox{0.5}{\includegraphics{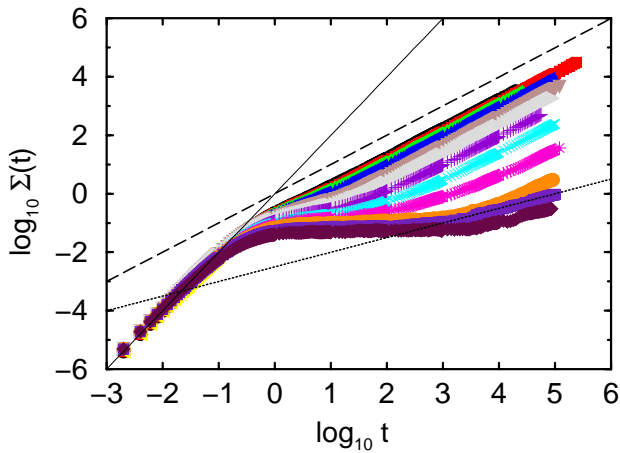}}
\end{center}
\vspace{-0.22in}
\caption{\label{msd} Mean-square displacement $\Sigma(t)$ versus time
$t$ for elastic hard-sphere systems with uniform $15\%$ polydispersity
over a range of packing fractions $\phi=0.50$ ($\circ$), $0.53$
($\triangle$), $0.54$ ($\lhd$), $0.55$ ($\triangledown$), $0.56$
($\rhd$), $0.57$ ($+$), $0.58$ (${\times}$), $0.59$ ($\ast$), $0.60$
($\bullet$), $0.61$ ($\Box$), and $0.62$ ($\diamond$) from top to
bottom.  The solid, dashed, and dotted lines have slopes $2$, $1$, and
$0.5$, respectively.  This data was obtained from studies by
M. Tokuyama and Y. Terada, and is similar to their results in
Ref.~\cite{tokuyama1}.}
\vspace{-0.22in}
\end{figure}

\begin{figure}
\begin{center}
\scalebox{0.5}{\includegraphics{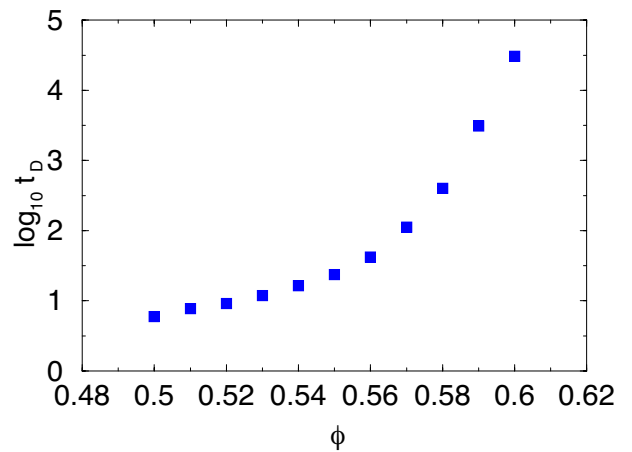}}
\end{center}
\vspace{-0.1in}
\caption{\label{relax} Structural relaxation time $t_D$ required
for the hard sphere systems in Fig.~\ref{msd} to reach the long-time
diffusive regime ($\Sigma(t) \sim D_L t$) as a function of packing fraction $\phi$.}
\vspace{-0.22in}
\end{figure}

Dense colloidal suspensions undergo a glass transition when they are
compressed to packing fractions $\phi$ approaching random close
packing (provided they are compressed rapidly or are sufficiently
polydisperse)~\cite{pusey}.  Random close-packed states are amorphous,
mechanically stable sphere packings with $\phi_{\rm rcp} \approx 0.64$
for monodisperse spheres~\cite{berryman,jam}.  In Fig.~\ref{msd}, we show the
mean-square displacement $\Sigma(t)$ (MSD) versus time $t$ over a
range of $\phi$ from $0.50$ to $0.62$ from molecular
dynamics (MD) simulations of polydisperse~\cite{rcp}, elastic hard
spheres with {\it ballistic} (not Brownian) short-time dynamics.  This
data was obtained from studies by M. Tokuyama and Y. Terada, and is
similar to results in Refs.~\cite{tokuyama1,angelani,doliwa}.  For
relatively dilute systems, the MSD crosses over from ballistic
($\Sigma(t) \sim t^2$) to diffusive ($\Sigma(t) \sim t$) when it
reaches $\approx 0.1 \sigma^2$, where $\sigma$ is the average particle
diameter.  The formation of a plateau in the MSD (for $\phi \gtrsim
0.57$) signals the onset of caging behavior, where particles are
trapped by neighboring particles that surround them.  The height and
length of the plateau characterize the cage size and the time over
which caging persists.  The appearance of the plateau and two-stage
relaxation in the MSD leads to dramatic increases in the structural
and stress relaxation times as shown in Fig.~\ref{relax}.  In this
figure, we demonstrate that the structural relaxation time $t_D$ (time
beyond which the MSD scales as ${\rm MSD} \sim D_L t)$ grows by nearly
four orders of magnitude over a small range in packing fraction.

Because of the rapid rise in relaxation times and the fact that dense
colloidal systems can only be equilibrated at packing fractions well
below random close-packing, it is difficult to accurately measure the
precise form of the divergence of the relaxation
times~\cite{zaccarelli}.  In particular, there is current vigorous
debate concerning the packing fraction at which complete dynamical
arrest occurs---is it before random close packing or does dynamic
arrest coincide with random close packing~\cite{chaikin}?  If it is
the former, it is possible that these systems undergo an ideal glass
transition to a static, but not mechanically stable state.  Further
open questions include determining the collective particle motions
that are responsible for subdiffusive behavior and the onset of
super-Arrhenius dynamics, which occur well below random close packing.

%%%%%%%%%%%%%%%%%
%%%%%%%%%%%%%%%%%

There have been a number of theoretical and computational studies
aimed at understanding slow dynamics in dense colloidal suspensions
and related glassy systems~\cite{cmaloney,amorphousmaterials, cohen,
dyre}.  These include the application of mode coupling theory to
colloidal systems~\cite{vanmegan1} and the development of
coarse-grained facilitated~\cite{jung} and kinetically constrained
lattice models~\cite{kinetic,lattice}.  Mode coupling theory has been
successful in predicting the form of the two-step relaxation of the
intermediate scattering function, but it predicts an
ergodicity-breaking transition well-above the experimentally
determined colloidal glass transition.  Related theories are able to
predict activated dynamics, but the location of the divergence in the
structural and stress relaxation times is still an input parameter,
not a prediction~\cite{schweizer}.  Models of dynamic facilitation, in
which mobile regions increase the probability that nearby regions will
also become mobile, are able to explain important aspects of dynamical
heterogeneities and non-Arrhenius relaxation times.  However, these
models have been implemented using either coarse-grained or lattice
descriptions, not particle-scale, continuum models.

%%%%%%%%%%%%%%%%%
%%%%%%%%%%%%%%%%%

Even though researchers have been able to visualize the motions of
colloidal particles in 3D using confocal microscopy for more than a
decade~\cite{weeks,weeks2}, an understanding of the particle-scale
origins of dynamical heterogeneities, cage formation and relaxation,
and structural rearrangements that give rise to subdiffusive behavior
is lacking.  Several factors have contributed to the slow progress.
First, it is well-known that it is difficult to predict dynamical
quantities from static structural properties.  Thus, even though one
can visualize all colloidal particles in 3D, it is difficult to
determine in advance which particles will move cooperatively.  Further,
it has proved difficult to identify and sample the rare transition
states that allow the system to move from one region in configuration
space to another.

\begin{figure*}
\includegraphics[width=1 \textwidth]{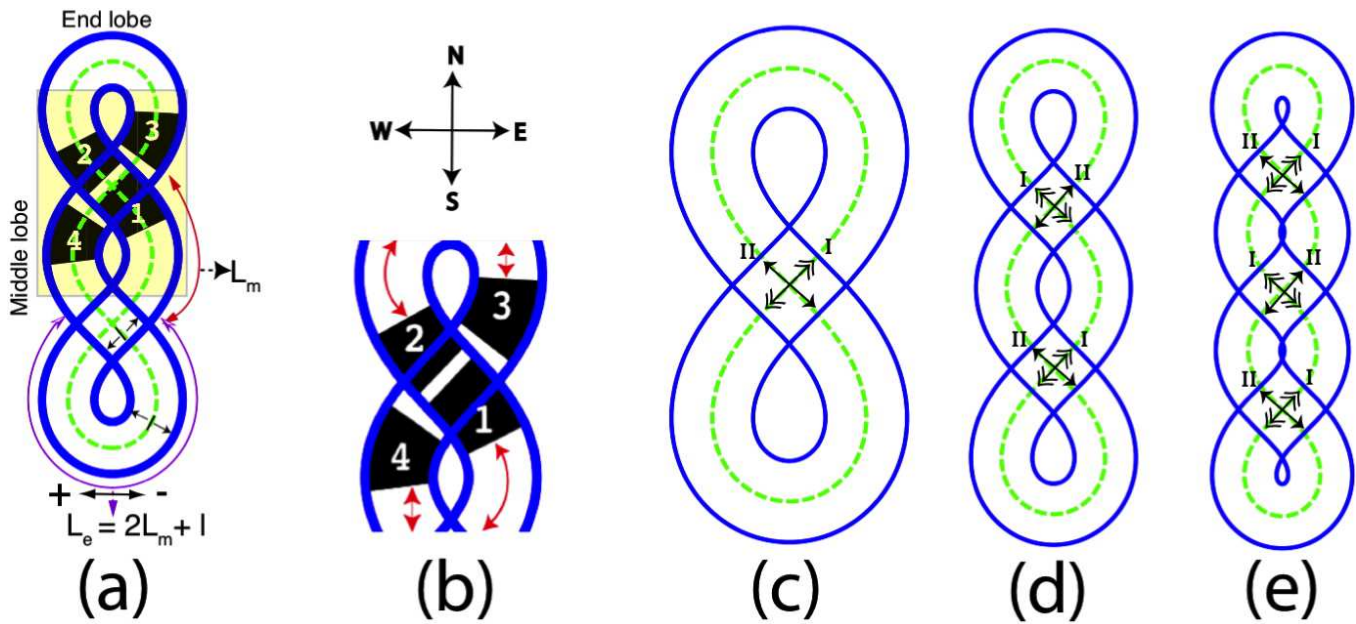}
\caption{\point{a} The Q1D channel consists of two different types of
lobes, middle and end, with lengths $L_m$ and $L_e = 2L_m + l$,
respectively. The width of the channel, $l$, is the same as the length
of the intersection. Particles move on the closed loop (green dashed
line) with an origin that is in the center of one of the end lobes.
The plus and minus directions are indicated. A close-up of the top
intersection is shown in \point{b}. When traversing the circuit,
particles move in direction $I$ for the first half of the circuit and
then in direction $II$ for the second half. Directions $I$ and $II$
alternate between the NE/SW and NW/SE directions for systems with
multiple junctions. Q1D channels are pictured with \point{c} $J = 1$,
(d) $2$, and (e) $3$ junctions.  The directions of motion I and II in
the junction are labeled.}
\label{system geometry}
\end{figure*}

We have developed a quasi-one-dimensional (Q1D) model, where hard rods
diffuse through a series of connected loops and junctions (or
intersections)~\cite{prasanta} as shown in Fig.~\ref{system geometry}.
There are a number of advantages for employing this model to explore
slow dynamics in colloidal and other glassy systems. First, this model
displays many features of glassy behavior including caging,
heterogeneous and collective dynamics, and a divergence of the
structural relaxation time $t_D$.  Second, the form of the divergence
of $t_D$ with increasing packing fraction can be determined
analytically.  Third, simulations and experiments of the colloidal
glass transition show evidence for quasi-one-dimensional behavior such
as correlated string-like motion of the fastest moving
particles~\cite{glotzer,glotzer1} and subdiffusive behavior with MSD $\sim
t^{0.5}$ that is characteristic of single-file diffusion in
quasi-one-dimensional systems~\cite{hahn,lutz}.  For example, possible
single-file subdiffusive behavior occurs in simulations of
polydisperse, elastic hard-sphere systems for
packing fractions $\phi = 0.61$ and $0.62$ as shown in Fig.~\ref{msd}.

In our previous studies of quasi-one dimensional models, we focused on
the `figure-8' system with a single junction (or
intersection)~\cite{prasanta} and $N$ hard rods. We found that the
structural relaxation time diverges as a power-law with increasing
packing fraction,
\begin{equation}
\label{taudeq}
t_D \sim (\phi^*-\phi)^{-\alpha},
\end{equation}
where $\alpha=N/2-1$ and $\phi^*=N/(N+4)$ is the packing fraction at which
kinetic arrest occurs. At kinetic arrest, a plateau in the MSD
persists for $t \rightarrow \infty$. Near $\phi^*$, the most likely
configurations are those with $N/2$ particles in both the top and
bottom end lobes, and no particles in the junction.  $t_D$ is
controlled by rare `junction-crossing' events, in which a particle
from the bottom (top) lobe, crosses the junction, enters the top
(bottom) lobe from one side of the junction, and another particle
exits the top (bottom) lobe and enters the bottom (top) lobe from the
other side of the junction. Thus, to undergo structural relaxation,
the system transitions from a relaxed configuration with half of the
particles in each lobe to a rare, squeezed configuration with an extra
particle in one of the lobes, and back to a relaxed configuration that
is similar to the initial one but with particle labels shifted forward
or backward by one.  The frequency $f$ of junction-crossing events is
determined by the probability $P_S$ for a squeezed configuration to
occur divided by the residence time that the system spends in the
squeezed configuration $\tau_r$,
\begin{equation}
\label{figure8eq}
f = \frac{P_S}{\tau_r}.
\end{equation} 
The structural relaxation time is the inverse of this frequency, and thus
$t_D=f^{-1}=\tau_r/P_S$. If we assume ergodicity, $P_S$ can be
calculated from configuration integrals, and for the figure-8 model,
$P_S \sim (\phi^*-\phi)^{N/2+1}$, where $N/2+1$ is the number of
particles in the squeezed lobe.  The residence time for the squeezed
configuration in the figure-8 model $\tau_r \sim (\phi^*-\phi)^2$
tends to zero in the limit $\phi^* - \phi \rightarrow 0$, and thus
$t_D \sim (\phi^*-\phi)^{-\alpha}$ as in Eq.~\ref{taudeq}.

\section{Model Description}
\label{definition}
\subsection{System geometry}
\label{Subsection-system-geometry}

We consider the collective dynamics of $\Nparticles$ non-overlapping
Brownian particles in a quasi-one-dimensional (Q1D) channel that forms
a closed loop with multiple intersections, as illustrated in Fig.\
\ref{system geometry}.  The particles move through the intersections
in mode \textit{I} in the first half of the circuit and
mode \textit{II} in the second half. For systems with multiple
intersections, modes $I$ and $II$ alternate between the
northeast/southwest (NE/SW) and northwest/southeast (NW/SE)
directions.  The particles can move in both the forward and backward
directions, but they cannot turn at the intersections.  Thus, to
switch the traffic mode at a given intersection, particles in one mode
must vacate the junction to allow particles in the other mode to
enter.

\begin{figure*}
\begin{center}
\scalebox{0.45}{\includegraphics{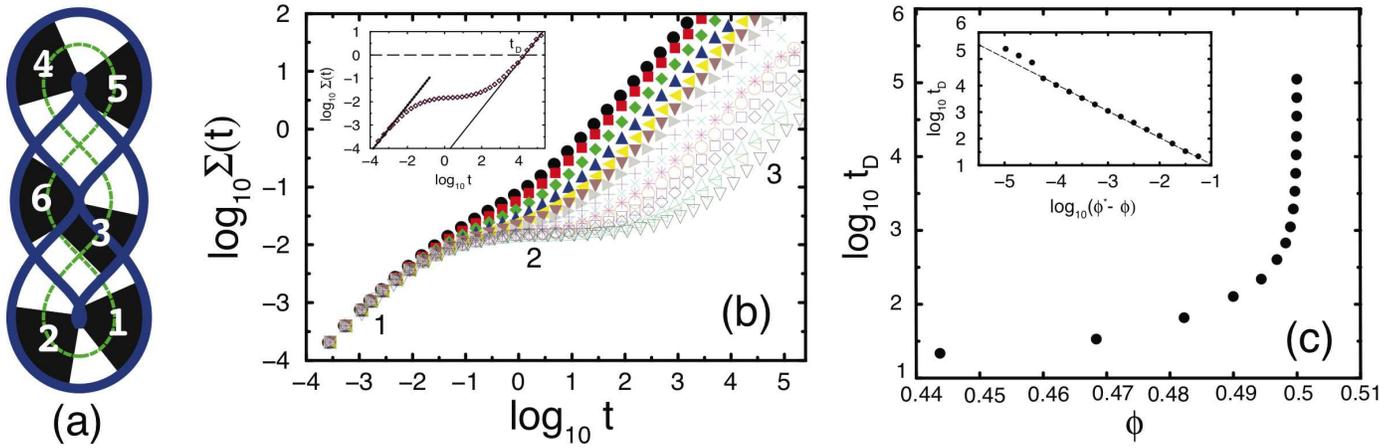}}
\end{center}
\vspace{-0.18in}
\caption{\label{fig:Figure-16msd}
\point a Q1D system with $N=6$, $J=2$, and $K=1$. (b) Mean-square
displacement (MSD) $\Sigma(t)$ versus time $t$ for the system in (a) from
$\phi=0.444$ (filled circles) to $ \approx 0.500$ (open downward triangles)
from left to right.  The numbers $1$, $2$, and $3$ indicate short-time
diffusive, plateau, and long-time diffusive behavior of the MSD,
respectively. (Inset) The time scale $t_D$ beyond which the system
displays diffusive behavior is obtained by setting $\Sigma(t_D)=1$
(long-dashed line).  The dotted and solid lines have slope $1$
corresponding to short- and long-time diffusive behavior,
respectively. \point b The timescale $t_D$ versus packing fraction $\phi$
for the Q1D model in (a). The slope of the long-dashed line in the
inset is $-1$. $t_D$ increases as a power law, $t_D \sim (\phi^* -
\phi)^{-1}$ with $\phi^* = 0.5$.
}
\vspace{0.22in}
\end{figure*}

The topology of the system is characterized by the number of junctions
$\junctions$.  Each channel has two end lobes, and for a given
$\junctions$ there are $2(\junctions-1)$ symmetric middle lobes.  The
channel geometry is described by three length parameters: the channel
width $\intersectionLength$ (which also determines the length of the
intersection), and the length of the end and middle lobes
$\lobeLengthE$ and $\lobeLengthM$.

To reduce the number of independent parameters
we focus on a model with
\begin{equation}
\label{end-lobe length}
\lobeLengthE=2\lobeLengthM+\intersectionLength.
\end{equation}
 We also assume that the particle size $\particleLength$ is equal to the
channel width,
\begin{equation}
\label{particle size and channel width are equal}
d=\intersectionLength.
\end{equation}
With these assumptions, exactly $K$ particles fit into a middle lobe
and $2K+1$ particles fit into an end lobe when $\lobeLengthM=K\intersectionLength$, where $K$ is an integer.
 
With the lengths of the middle and end lobes
related by Eq.\ \eqref{end-lobe length}, the total length of the channel  is
\begin{equation}
\label{channel length}
\channelLength=2(\junctions+1)(\lobeLengthM+\intersectionLength),
\end{equation}
where the length of each intersection is counted both in the NE and NW
directions.  The packing fraction of the particles in the
channel is given by
\begin{equation}
\label{packing fraction}
\phi=\frac{\Nparticles \intersectionLength}{\channelLength}.
\end{equation}

In our intersecting-channel model, particles moving through an
intersection in one direction block the motion of particles in the
perpendicular direction.  Thus, at high packing fractions the system
undergoes kinetic arrest.  In a kinetically arrested (KA)
configuration, the particles can perform local movements,
but the system cannot undergo collective rearrangements that lead to
diffusive motion at long time scales.
In this work, we analyze the slow dynamics of Q1D systems as   
the packing fraction is increased by changing
the lobe length $\lobeLengthM$ at  constant particle number
$\Nparticles$.

\subsection{System dynamics}
\label{Brownian dynamics}

In our model, each particle undergoes Q1D Brownian
motion~\cite{orstein} along the channel length.  This Brownian motion
is implemented numerically using a Monte Carlo algorithm
~\cite{montecarlosimulation,malvin,frenkel} with random
single-particle moves and the step size chosen from a Gaussian
distribution~\cite{nrc}.  
The standard-deviation of the Brownian step distribution $\sigma$ is
chosen small enough to accurately represent Brownian dynamics of
non-overlapping particles with short time diffusion coefficient
$D_s\propto\sigma^2$.  At low packing fractions we use
$\sigma=0.1\Delta$, where
$\Delta=(\channelLength-\Nparticles\intersectionLength )/\Nparticles$
is the average gap size between particles.  For large packing
fraction, we reduced the standard deviation to $\sigma \propto (L_e -
(2K+1)l)/N_e$, where $N_e$ is the number of particles in the most
occupied end lobe, to ensure that rare configurations are sampled.

\begin{figure*}
\includegraphics[width=0.95\textwidth]{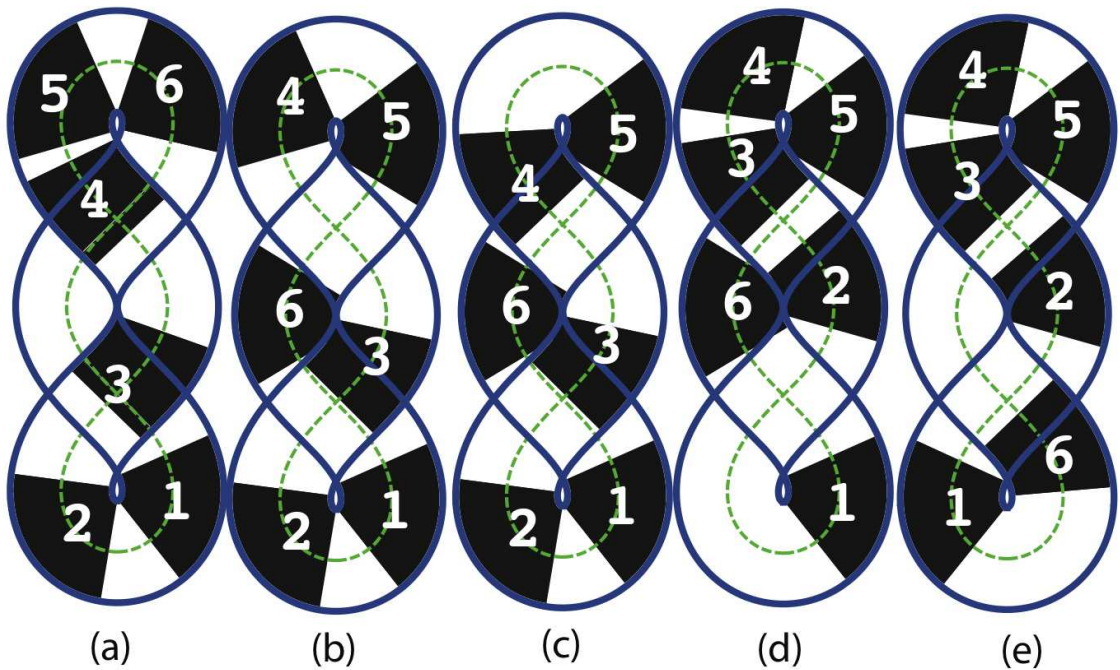}
\caption{Illustration of the bottleneck event that causes slow
dynamics in the Q1D system with $N=6$, $J=2$, $K=1$, and $M=2$
pictured in Fig.\ \ref{fig:Figure-16msd}. For a rearrangement event to
occur, particle $6$ must migrate into the middle lobe (\point{a} and
\point{b}), reside there until other particles ($2$ and $3$) pass from
the upper to the lower part of the channel (\point{c} and \point{d}), and then pass through the
lower intersection into the lower end lobe \point{e}.}
\label{example of bottleneck}
\end{figure*}

\subsection{Close-packing and kinetic arrest}
\label{Close-packing-jamming-kinetic-arrest}

It is important to emphasize that KA states are distinct from
close-packed configurations.  In a close-packed
configuration, some or all particles in the system cannot move, and
the system size $\lobeLengthM$ cannot be reduced in a continuous manner without
creating particle overlap.  In KA states, local particle motions are
possible (and $\lobeLengthM$ can be reduced), but the particles are blocked at
the intersections, and no particles are able to complete a full circuit 
around the channel. 

There are two possible types of behavior for systems with middle lobe
lengths that are slightly above the critical value
$\lobeLengthM=\lobeLengthMKA$ for kinetic arrest.  After passing
through the geometrical bottleneck associated with kinetic arrest,
the system either arrives at an unconstrained state where the
particles diffuse around the circuit on a timescale of the order of
$\tauD=\channelLength^2/D_s$, or remains constrained by a sequence of
bottlenecks that need to be cleared to complete a 
circuit.

We are interested here in the kinetic arrest of the second kind, where
not only the initial escape from the nearly KA state occurs on a
divergent timescale, but the timescale for the subsequent long-time
diffusive behavior also diverges at $\lobeLengthM=\lobeLengthMKA$.
In what follows, the term kinetic arrest refers only to the second-kind
behavior.

\subsection{Critical dynamics near kinetic-arrest threshold}
\label{Critical behavior near kinetic arrest}

The characteristic dynamics in the system for $\lobeLengthM$ approaching the
critical value $L_m^*$ is illustrated in Fig.\ \ref{fig:Figure-16msd}.
As depicted in Fig.\ \ref{fig:Figure-16msd} \point{a}, the system has
$\junctions=2$ junctions and contains $\Nparticles=6$ particles.
Fig.\ \ref{fig:Figure-16msd} \point{b} shows the mean-square
displacement (MSD) $\Sigma(t)$ of the particles versus time $t$, and
Fig.\ \ref{fig:Figure-16msd} \point{c} depicts the time the system
needs to reach the long-time diffusive regime (where $\Sigma(t) \sim
D_L t$) for $L_m$ slightly above the kinetic-arrest threshold
$\lobeLengthMKA=\intersectionLength$ or packing fraction slightly
below the critical packing fraction ($\phiKA=0.5$).

The results in Fig.\ \ref{fig:Figure-16msd} \point{b} show that near
the KA threshold $\phi=\phiKA$ the Q1D model displays slow dynamics
that resembles the dynamics observed in glass-forming systems.  We
find three dynamical regimes: \point{a} short-time diffusion,
\point{b} the formation of a plateau, where $\Sigma(t)$ remains
nearly constant, and \point{c} long-time diffusion.  As shown in Fig.\
\ref{fig:Figure-16msd} \point{c}, the long-time diffusive motion is
arrested at $\phi=\phiKA$.  A cursory examination of the system
depicted in Fig.\ \ref{fig:Figure-16msd} \point{a} is insufficient to
determine the mechanism that causes the rapid growth of the timescale required to reach the 
long-time diffusive regime as shown in Fig.\ \ref{fig:Figure-16msd} \point{b}
and \point{c}.  

A detailed analysis (cf.\ Sec.\ \ref{Discrete microstates and
microstate network}) reveals that the bottleneck causing the slow
diffusion corresponds to the dynamical event depicted in Fig.\
\ref{example of bottleneck}. During this event one of the particles
(particle $6$ in the example considered) needs to migrate into the
middle lobe and reside there until other particles pass from the upper
to the lower part of the channel.  In the limit
$L_m \rightarrow L_m^* = l$, the particle residing in the
middle lobe does not have enough room to move, which results in a low
probability of this squeezed configuration and implies that the
corresponding bottleneck-crossing event is rare.
 
\subsection{Critical packing fractions}
\label{Critical packing fractions and lob lengths}

Our numerical simulations indicate that kinetic arrest with divergent
timescales required to reach the long-time diffusive regime occurs for
critical lobe lengths equal to integer multiples of the intersection
or particle length,
\begin{equation}
\label{integer lobe length}
\lobeLengthMKA=\Klobe\intersectionLength.
\end{equation}
Since each of the $\junctions$ junctions can be filled by at most
a single particle, the maximal number of particles in a system with
$\lobeLengthM=\lobeLengthMKA$ is
\begin{equation}
\label{clos packing N}
\Ncp=2(\junctions+1)(\Klobe+1)-\junctions.
\end{equation}
The corresponding close-packing fraction is
\begin{equation}
\label{close packing phi}
\phicp(\junctions,\Klobe)=\frac{\Ncp}{\Ncp+\junctions}.
\end{equation}

According to our analysis presented in
Sec.~\ref{Discrete microstates and
microstate network}, a system near the KA threshold
\eqref{integer lobe length} requires at least two particle-size
vacancies to allow long-time diffusive motion.  One vacancy is needed
to empty an intersection, and the other to allow a particle moving in
the other direction to completely cross the intersection.  

In fact, a system with $J$ junctions and lobe occupation number $\Klobe$
exhibits critical scaling of $t_D$ in the presence of $2+\Mvoids$ voids,
i.e., for
\begin{equation}
\label{number of particles at KA}
\Nparticles=\Ncp-2-\Mvoids
\end{equation}
particles, where
\begin{equation}
\label{range of }
0\le\Mvoids\le\MvoidsMax
\end{equation}
and 
\begin{equation}
\label{m_max}
\MvoidsMax = 2(J + 1)K
\end{equation}
is the maximum number of particle size voids in the system such that
when $M\rightarrow\MvoidsMax$ the system still undergoes kinetic
arrest. If one more particle size void is added to the system (or
conversely a particle is taken out), the system no longer requires a
`squeezed', bottleneck configuration to relax. Thus, the packing
fraction for kinetic arrest is
\begin{equation}
\label{KA packing fractions}
\phiKA(\junctions,\Klobe,\Mvoids)=\frac{\Ncp-\Mvoids-2}{\Ncp+\junctions}.
\end{equation}

\begin{figure}
\scalebox{0.85} {\includegraphics[width=0.3\textwidth]{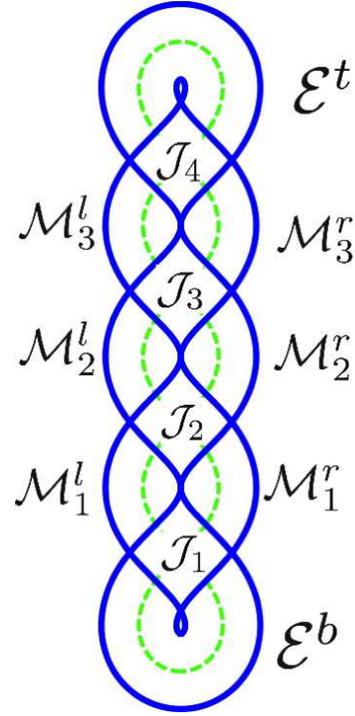}}
\caption{Each Q1D configuration can be mapped to one of the discrete
microstates $S = \{ \mathcal{E}^b \mathcal{J}_1 \mathcal{M}^r_1
\mathcal{M}^l_1 \ldots
\mathcal{J}_{J-1} \mathcal{M}^r_{J-1} \mathcal{M}^l_{J-1} \mathcal{J}_J \mathcal{E}^t\}$,
which is a set of integers that represents the occupancy of the lobes
and intersections of the system.  The integer $\mathcal{E}^b$
($\mathcal{E}^t$) is the number of particles in the bottom (top) end lobe and
$\mathcal{M}_i^r$ and $\mathcal{M}_i^l$ give the numbers of particles in the
$i$th right and left middle lobes. The integer $\mathcal{J}_i$ represents the
state of junction $i$ defined by Eq.~\ref{junction variable}.
}
\label{state-notation-figure}
\end{figure}

\begin{figure*}
\includegraphics[width=0.9 \textwidth]{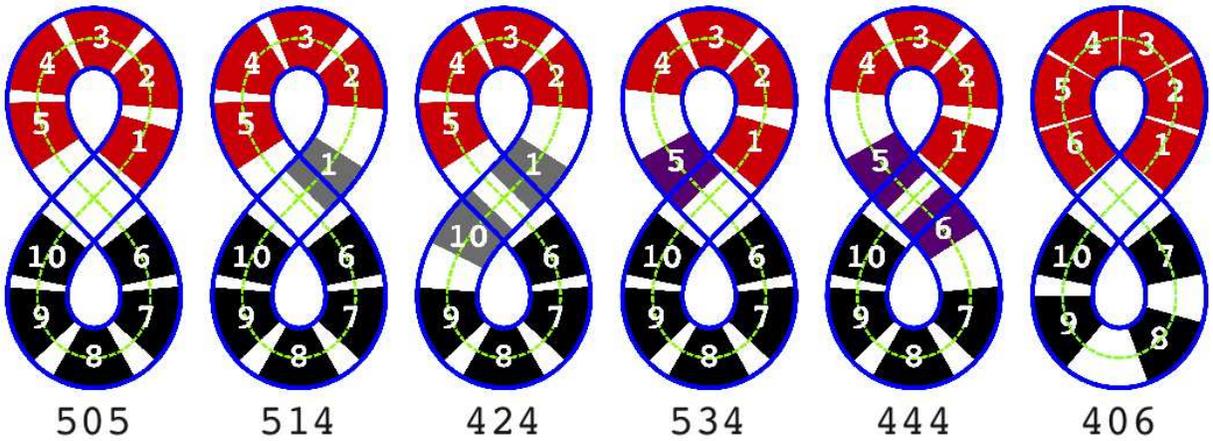}
\caption{(Color online) Illustration of microstates $505$, $514$, $424$, $534$,
$444$, and $406$ for Q1D systems with $N=10$ and $J=1$. Red and black
shaded particles occupy the top and bottom end lobes, respectively.
Gray and purple shaded particles occupy the junction (or intersection)
in directions $I$ and $II$, respectively.}
\label{microstateexample}
\end{figure*}

\section{Discrete microstates and construction of microstate network}
\label{Discrete microstates and microstate network}

To describe the structural relaxation mechanisms in the Q1D-channel
model, it is convenient to map all of the configurations of the system
onto a set of discrete microstates.  The microstates correspond to
configuration-space regions defined by: (\textit{i}) the number of
particles residing in each lobe; and (\textit{ii}) the number of
particles present in each intersection and their direction of motion.
Such a discrete mapping allows us to employ graphical techniques to
identify bottleneck states that control the slow dynamics of the system.

\subsection{Definition of Microstates}
\label{statedef}

We represent each discrete microstate by the occupancy variable
\begin{equation}
\label{occupancy variable}
\occupancyVar=\{
  \mathcal{E}^b \mathcal{J}_1,\mathcal{M}^r_1 ,\mathcal{M}^l_1 \ldots 
  ,\mathcal{J}_{J-1},\mathcal{M}^r_{J-1} ,\mathcal{M}^l_{J-1} ,\mathcal{J}_J 
  ,\mathcal{E}^t\},
\end{equation}
which is the set of integers that represents the states of lobes and
intersections (as illustrated in Fig.\ \ref{state-notation-figure}).
The integer $\mathcal{E}^b$ ($\mathcal{E}^t$) is the number of
particles in the bottom (top) end lobe and $\mathcal{M}^r_i$
($\mathcal{M}^l_i$), $i=1\ldots J$, is the number of particles in the
$i$th right (left) middle lobe.  A particle is assumed to reside in
the lobe if its entire length is contained within the lobe length.

If any portion of a particle enters an intersection, the particle is
assigned to this intersection.  Since the particle length is the same
as the intersection length (Eq.\ \eqref{particle size and channel
width are equal}), the maximal number of particles that can reside in
a given intersection is two.  The state of intersection $i$, which is
occupied by $k_i$ particles, is given by
\begin{equation}
\label{junction variable}
\mathcal{J}_{i}= 2(1-\delta_{q_i 0})(1-\delta_{k_i 0})+k_i,
\end{equation}
where $q_i=0$ and $1$ for directions of motion \textit{I} and
\textit{II} (as defined in Fig.\ \ref{system geometry}).  Several
examples of microstates and their corresponding occupancy variables
\occupancyVar~are illustrated in Fig.\ \ref{microstateexample} for a
figure-8 model with $N=10$ and $J=1$.  The examples show all five
states of the intersection, $\mathcal{J}_1=0,\ldots,4$, which is the 
middle integer in the microstate label.

\begin{figure*}
\begin{center}
\scalebox{0.3}{\includegraphics{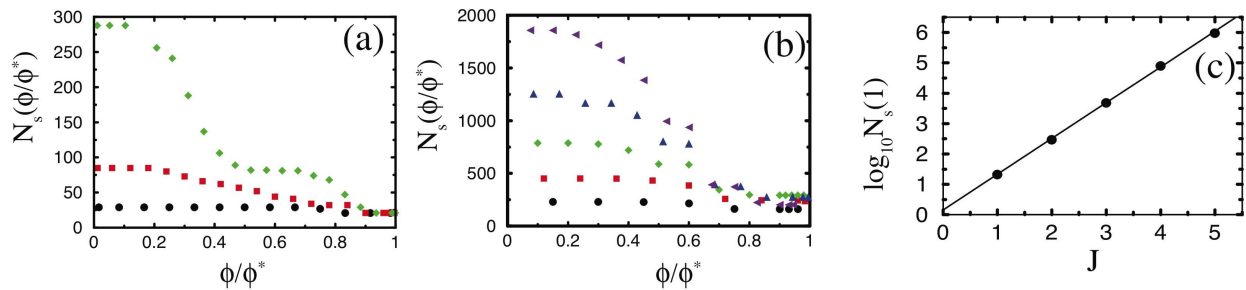}}
\end{center}
\vspace{-0.18in}
\caption{The number of microstates $N_S$ as a
function of $\phi/\phi^{*}$ for a Q1D model with \point{a}
$J=1$, $N=6$ (circles), $20$ (squares), $100$ (diamonds), and (b)
$J=2$, $K=1$, and $N=4$ ($M=6$; circles), $5$ ($M=5$; squares), $6$
($M=4$; diamonds), $7$ ($M=3$; upward triangles), and $8$ ($M=2$;
leftward triangles). \point {c} Number of microstates at $\phi=
\phi^*$ for a Q1D system with $K=1$, $M=J$, and $N=N_{cp} - 2 - M$.}
\label{stateVsPhi}
\vspace{0.22in}
\end{figure*}

The number of microstates $N_S(\phi)$ allowed by the excluded-volume
constraints is maximal at $\phi\rightarrow 0$ and decreases as the 
packing fraction approaches $\phiKA$. In Fig.~\ref{stateVsPhi} (a)
and (b) we show that for a fixed topology, the number of microstates
$N_S(\phiKA)$ at kinetic arrest does not depend or only weakly
depends on the number of particles.  In contrast, the number of microstates
grows exponentially when the number of particles $N$ and intersections
$J$ is increased simultaneously, as depicted in Fig.~\ref{stateVsPhi}
(c).

\subsection{Classification of states: Squeezed and trapped microstates}
\label{Classification of states: squeezed and trapped microstates}

As discussed in Secs.\ \ref{intro} and \ref{Critical behavior near
  kinetic arrest}, the system dynamics near the KA threshold
$\phi\to\phiKA(\junctions,\Klobe,\Mvoids)$ is controlled by
low-probability bottleneck microstates through which the system must pass
to continuously move the particles around the channel.  The
bottleneck microstates occur with low probability, $P_S$, because they
correspond to a vanishingly small portion of the configuration space
when the system approaches kinetic arrest.  Thus, the infrequent sampling of 
the bottleneck microstates results in the rapidly growing timescale 
required to reach the long-time diffusive regime as $\phi \rightarrow \phi^*$. 

%%%%%%%%%%%%%%%%%

\subsubsection{Types of squeezed states}

To facilitate the identification of the bottleneck states and analysis
of the scaling behavior of the structural relaxation time $t_D$ near
the KA packing fraction $\phiKA(\junctions,\Klobe,\Mvoids)$, we
decompose all microstates into two main categories: the sets of
unsqueezed ($\unsqueezedSet$) and squeezed ($\squeezedSet$)
microstates.  For unsqueezed microstates, none of the lobes is
completely filled with particles at
$\phi=\phiKA(\junctions,\Klobe,\Mvoids)$.  In contrast, for squeezed
microstates at least one lobe becomes completely filled (i.e. squeezed or
compressed) when $\phi\to\phiKA(\junctions,\Klobe,\Mvoids)$.  Hence,
the configuration-space volume corresponding to squeezed microstates
vanishes when $\phi = \phi^*$, whereas the
volume corresponding to unsqueezed microstates remains nonzero.

\begin{figure*}
\begin{center}
\scalebox{0.5} {\includegraphics{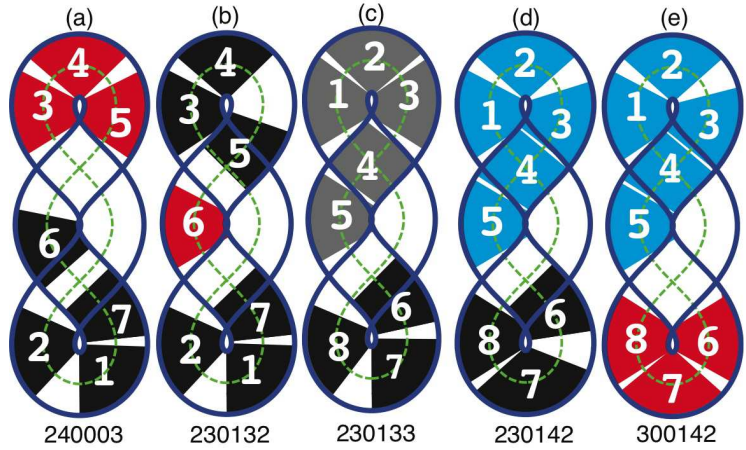}}
\end{center}
\caption{(Color online) Illustration of the types of compressed regions (CRs) for a
Q1D system with $N=7$, $J=2$, $K=1$, and $M=1$: simple CRs (red) with
compressed (a) top end and (b) left middle lobes; (c) a composite CR
(gray) with compressed top end and left middle lobes with particle $4$
moving in direction $I$; (d) redistributed CR (blue) with particles
$3$ and $4$ occupying the top intersection moving in direction $I$,
and (e) a composite squeezed state cluster composed of the redistributed 
CR (blue) from (d) and simple CR (red) in the bottom end lobe.}
\label{CR}
\end{figure*}

A squeezed microstate can contain one or more compressed regions
(CRs).  For the channel geometry described in Sec.\
\ref{Subsection-system-geometry}, there are three types of CRs. First, a
\textit{simple} CR consists of a single compressed lobe, e.g. a
compressed top end or left middle lobe as shown in Fig.~\ref{CR} (a)
and (b). According to Eq.~\ref{end-lobe length} and the notation
introduced in Sec.\ \ref{Critical packing fractions and lob lengths},
compressed middle and end lobes contain $\Nc=\Klobe$ and
$\Nc=2\Klobe+1$ particles, respectively.

Second, a \textit{composite} CR (shown in Fig.~\ref{CR} (c)) is a
contiguous region that consists of compressed lobes and intersections
that connect them. Each connecting intersection contains a single
particle that is moving in a direction that will connect the lobes
(without causing a turn at the intersection).  The number of particles in a
composite CR that includes $\km$ compressed middle lobes and $\ke$
compressed end lobes is
\begin{equation}
\label{expression for Nc}
\Nc=\km(K+1)+2\ke(K+1)-1.
\end{equation}

A \textit{redistributed} CR, as shown in Fig.~\ref{CR} (d), is a region that
can be obtained from a composite CR by moving some particles from the
compressed lobes to the adjacent connecting intersections.  The number
of particles in a redistributed CR is the same as the number of
particles in the corresponding composite CR given by
Eq.\ \eqref{expression for Nc}.

We define a \textit{simple squeezed microstate} to be one 
that contains only simple CRs.  Squeezed microstates that are
not simple form a \textit{composite squeezed-microstate cluster}, such as 
that shown in Fig.~\ref{CR} (e), which is
a set that contains (\textit{i}) a given squeezed microstate
$\squeezedSet$ that includes only simple and composite CRs and
(\textit{ii}) all microstates that can be obtained from $\squeezedSet$ by
replacing composite CRs with the corresponding redistributed CRs.

The particles contained in CRs of a given squeezed microstate are termed
\textit{compressed particles}.  The total number of compressed
particles in a squeezed microstate that has $k$ CRs is
\begin{equation}
\label{total number of squeezed particles}
\Nc=\sum_{i=1}^k\Nci{i},
\end{equation}
where $\Nci{i}$ is the number of particles in the $i$th CR.  As will
be discussed in Secs.\ \ref{Untrapped, trapped, and KA squeezed
microstates} and \ref{Relaxation timescales}, the number of squeezed
particles $\Nc$, combined with the effects of trapping on the ends of
the CRs, determine the scaling (with $\phi^*-\phi$) of the frequency
$f$ with which a compressed microstate (or microstate cluster) is
sampled as the system evolves at long times.

\subsubsection{Untrapped, trapped, and KA squeezed microstates}
\label{Untrapped, trapped, and KA squeezed microstates}

To estimate how long, on average, the system resides in a given
squeezed microstate, we introduce the concept of {\it microstate
trapping}.  To this end, we first establish three types of boundaries
of a CR as shown in Fig.~\ref{ends}.  The boundary (i.e., the
intersection that terminates the first or last lobe in a CR) is:
\begin{itemize}

\item
\textit{free} if the terminal intersection is empty or the
direction of particle motion in this intersection is along the line passing through the
compressed terminal lobe;

\item \textit{trapping} if the direction of particle motion in the terminal intersection is
  orthogonal to the compressed lobe, and the intersection is not a
  part of a compressed region;

\item \textit{kinetically arresting} if the direction of particle motion in the terminal
  intersection is orthogonal to the compressed lobe, and the
  intersection is a part of a compressed region.

\end{itemize}

A squeezed microstate (consisting of one or more CRs) is \textit{untrapped}
if at least one CR end is free.  In a \textit{trapped} state there are
no free CR ends, but at least one end is trapping.  If all CR ends are
kinetically arresting, the microstate is KA, and the evolution is
constrained to this microstate or the associated microstate cluster.
Since the microstate
occupancy variable \eqref{junction variable} specifies the number of
particles in each lobe as well as both the number of particles and the
direction of motion for each intersection, the untrapped, trapped, and KA
CRs can be identified by analyzing the sequence of integers in
$\occupancyVar$.
As further discussed in Sec.\ \ref{Relaxation timescales}, trapped,
squeezed microstates relax more slowly than untrapped, squeezed microstates,
which influences the frequency of rare microstate sampling.

\begin{figure}
\scalebox{1} {\includegraphics[width=0.3\textwidth]{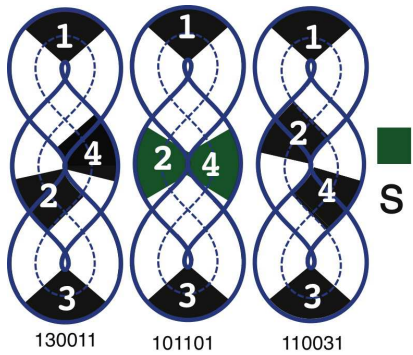}}
\caption{(Color online) Illustration of semitrapped middle lobes (S: green) in microstate
$101101$. When the system transitions to microstate $130011$, particle 
$4$ is prevented from moving downward and particle $2$ is prevented from 
moving upward.  A similar effect occurs when the system transitions to 
$110031$.}
\label{semi}
\end{figure}

\begin{figure*}
\begin{center}
\scalebox{0.5} {\includegraphics{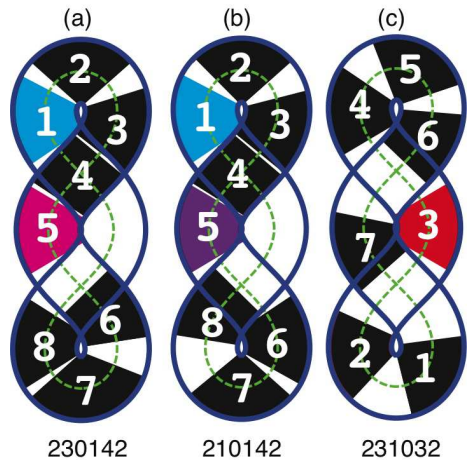}}
\end{center}
\caption{(Color online) Illustration of the three types of ends of
compressed regions (CRs) using a Q1D system with $N=7$, $J=2$, $K=1$,
and $M=2$. (a) The redistributed CR formed by particles $1$, $2$, $3$,
$4$, and $5$ has kinetically arrested (particle $1$; blue) and free
(particle $5$; pink) ends. (b) The redistributed CR formed by particles
$1$, $2$, $3$, $4$, and $5$ has kinetically arrested (particle $1$;
blue) and trapped (particle $5$; violet) ends. (c) The simple CR
formed by particle $3$ (red) has two trapping ends (particles $6$ and $7$).}
\label{ends}
\end{figure*}

In addition to the three basic types of squeezed microstates described
above, we consider a special case of a compressed pair of left and
right middle lobes for a system with lobe size $\Klobe=1$ in the
middle panel of Fig.\ \ref{semi}.  As discussed in Sec.\
\ref{Relaxation timescales}, the crossing frequency of such microstates is not
controlled by particle dynamics within the CRs, but by particle
motion in the neighborhood of the CRs. See the left and right panels of Fig.\
\ref{semi}.  We will refer to these systems as
\textit{semitrapped}.

\subsubsection{Relaxation timescales}
\label{Relaxation timescales}

By the arguments leading to Eq.\ \eqref{figure8eq}, the frequency of
crossing a bottleneck microstate $\occupancyVar$ (or an associated microstate
cluster) depends on the microstate probability $P_S$ and on the time $\tau_r$
the system spends in microstate (cluster) $\occupancyVar$ during a crossing
event.  Due to system ergodicity, the probability $P_S$ is proportional
to the fraction of the configurational-space volume the microstate
(cluster) occupies.  For a squeezed microstate (cluster) with $\Nc$
compressed particles, the probability scales as
\begin{equation}
\label{probability scaling for squeezed state}
P_S\sim\Omega \sim(\phiKA-\phi)^\Nc,
\end{equation}
where $\Omega$ is the configurational-space region occupied by the bottleneck
microstate (cluster). 

The analysis described in our previous study \cite{prasanta} shows
that the residence time $\tau_r$ for an untrapped, squeezed microstate
scales as
\begin{equation}
\label{residence time untrapped}
\tau_r\sim(\phiKA-\phi)^2.
\end{equation}
The residence time \eqref{residence time untrapped} is the timescale
for an end particle in a CR to diffuse a distance proportional to
$\phiKA-\phi$ to the CR border.  In contrast, for a trapped simple CR,
\begin{equation}
\label{residence time trapped}
\tau_r \sim O(1),
\end{equation}
because the particles blocking the intersections need to diffuse an
$O(1)$ distance to release the trapped particles.  A trapped composite
CR can relax to an associated redistributed CR on the fast timescale
\eqref{residence time untrapped}; however, the system remains in the
cluster of trapped microstates for the longer time interval
\eqref{residence time trapped}.

The scaling of the frequency (Eq. \eqref{figure8eq}) for crossing a
microstate corresponding to a simple or composite CR near a KA
transition is obtained by combining \eqref{probability scaling for
squeezed state} with \eqref{residence time untrapped} for untrapped
microstates and with \eqref{residence time trapped} for trapped microstates.
Thus, the bottleneck crossing frequency is 
\begin{equation}
\label{frequency of state crossing general}
f\sim(\phiKA-\phi)^\crossingExponent,
\end{equation}
where
\begin{subequations}
\label{crossing exponent}
\begin{equation}
\crossingExponent=\Nc-2
\label{crossing exponent U}
\end{equation}
and
\begin{equation}
\crossingExponent=\Nc
\label{crossing exponent T}
\end{equation}
are the crossing-frequency exponents for untrapped states with $\Nc>2$
and for trapped states, respectively.  

For a semitrapped microstate illustrated in Fig.\ \ref{semi}
there is no geometrical trapping (i.e., the particles are free to
leave the CR).  Since $\Nc=2$, Eq.\ \eqref{crossing exponent U}
predicts $\crossingExponent=0$ in this case.  However, our numerical
simulations indicate that, instead, the crossing frequency scales with
the exponent
\begin{equation}
\crossingExponent=1.
\label{crossing exponent S}
\end{equation}
\end{subequations}
This anomalous behavior indicates that the crossing frequency is not
controlled by the CR itself, but by particle dynamics in its
neighborhood during the approach to and subsequent separation from the CR.

Relation \eqref{crossing exponent U} for untrapped CRs can be derived
using an alternative first-passage time argument.  Accordingly, we
treat the boundary of an untrapped CR in $\Nc$--dimensional
configuration space as an absorbing surface, and consider a stationary
probability distribution $\rho$ that tends to the constant equilibrium
value at infinity.  By solving the $\Nc$--dimensional Laplace equation
for this boundary-value problem, we find that the perturbation
$\delta\rho$ of the probability distribution due to the presence of
the absorbing boundary scales as
\begin{equation}
\label{first passage time argument delta rho}
\delta\rho\sim(R/r)^{\Nc-2},
\end{equation}
where $R\sim\phiKA-\phi$ is the characteristic dimension of the CR,
and $r$ is the distance from the CR region.  Integrating the
corresponding probability flux density 
\begin{equation}
\label{flux density}
j_\rho\sim r^{-1}(R/r)^{\Nc-2}
\end{equation}
over the $(\Nc-1)$--dimensional CR surface yields
\begin{equation}
\label{total flux}
J_\rho\sim R^{\Nc-2},
\end{equation}
consistent with Eq.\ \eqref{crossing exponent U}.  

We note that the above argument does not apply to a semitrapped CR,
because the solution of the corresponding 2D Laplace equation for the
probability density $\rho$ diverges logarithmically at infinity.  This
logarithmic divergence may suggest that the crossing frequency $f$
decays logarithmically when packing fraction $\phi$ tends to the KA
value; however, our numerical simulations yield the power-law behavior
\eqref{crossing exponent S}.  Resolving this discrepancy requires
further study.

\begin{figure*}
\begin{center}
\scalebox{2} {\includegraphics{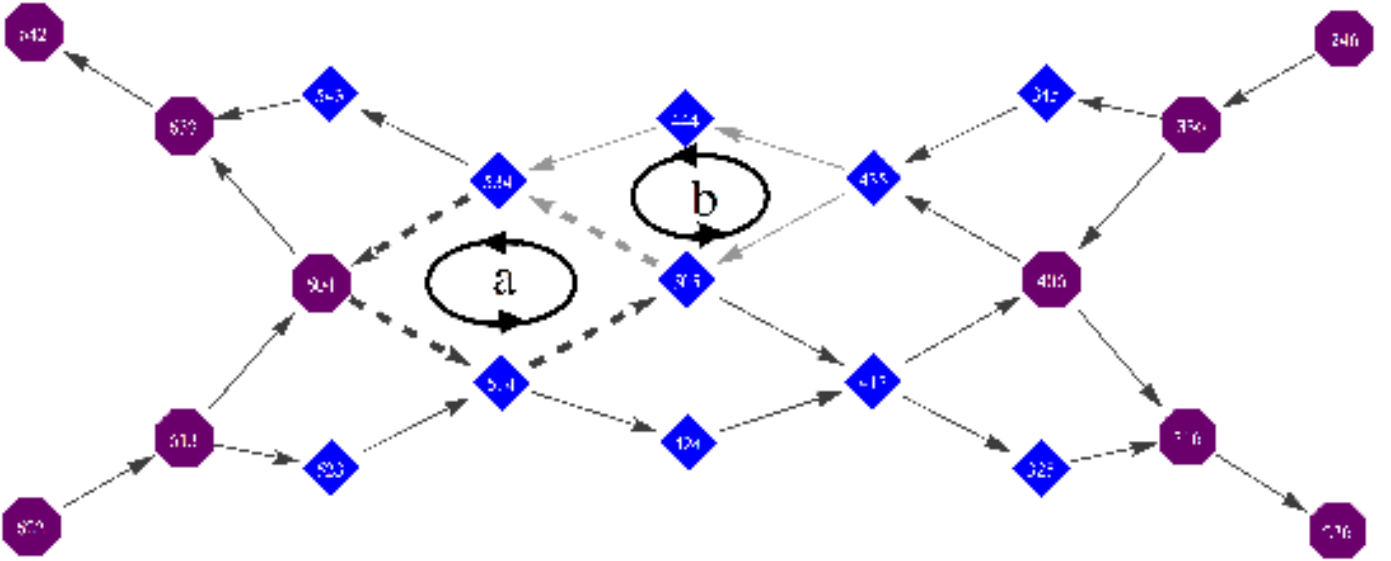}}
\end{center}
\caption{\label{fig-8-network} Directed network containing $21$
interconnected microstates for the Q1D model with $N=10$ and
$J=1$. The octagons and diamonds indicate squeezed and unsqueezed
microstates, respectively.  The (a) loop (formed by the microstates
connected by dashed arrows) possesses nonzero winding number $W = 4$,
whereas the (b) loop (formed by the microstates connected by the gray
arrows) has $W=0$.  }
\end{figure*}

\subsection{Diffusion through the microstate network}
\label{Diffusion of microstate network}

For small systems such as the figure-8 with $J=1$ considered in Ref.\
\cite{prasanta}, the bottleneck microstates can be determined by
inspection.  However, when the number of particles and intersections is
increased, the number of microstates grows exponentially (Fig.~\ref{stateVsPhi}
(c)), and the system becomes rapidly too complex for a simple analysis.
To facilitate an automated analysis, we represent the system evolution
as a diffusive process on a network of connected microstates.  Key
features of the network are determined using graph-theoretical techniques.

\subsubsection{States, transitions, and graphs representing the microstate network}
\label{Transition between states}

\begin{figure*}
\includegraphics*[width=1\textwidth]{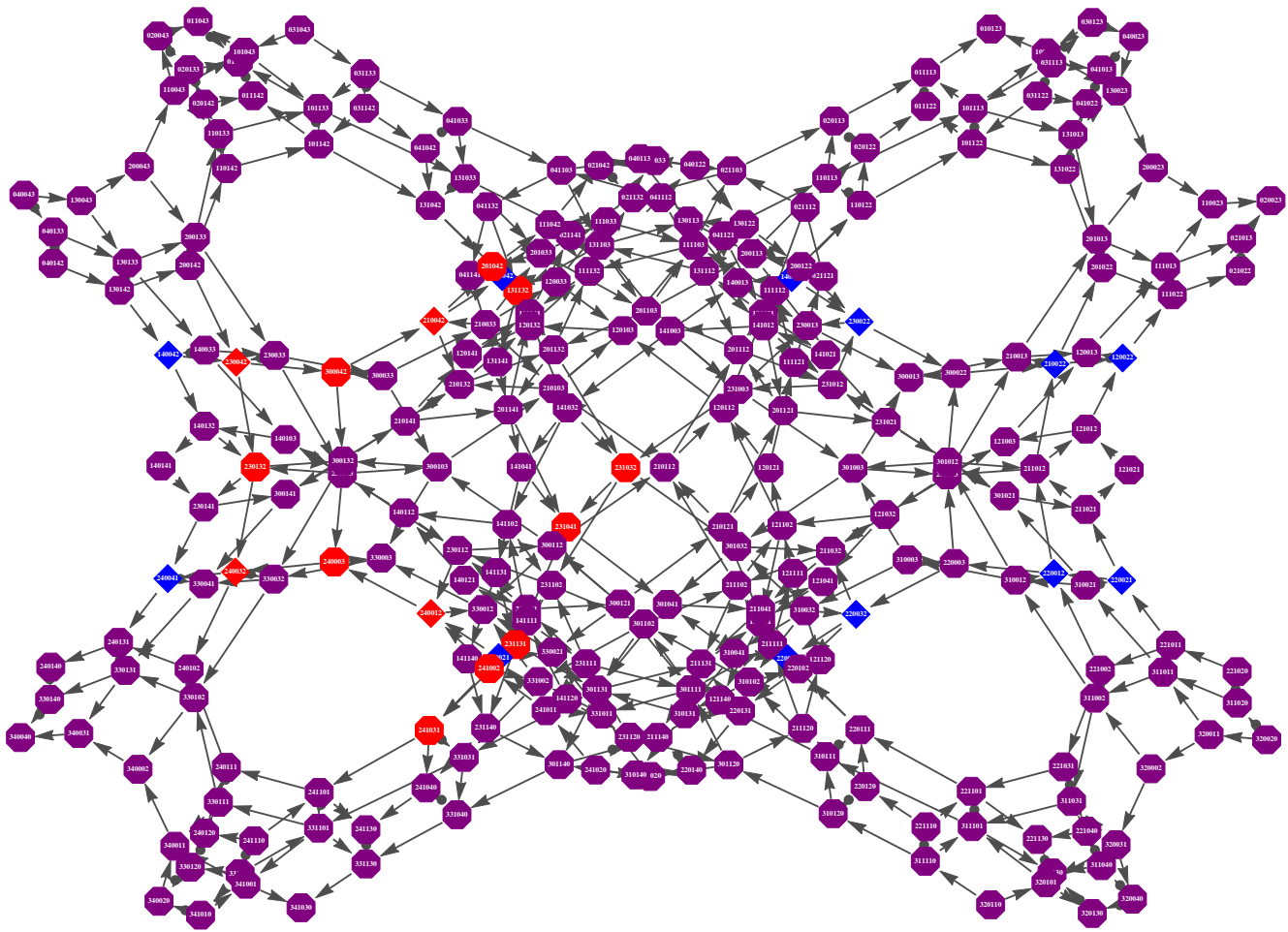}
\caption{(Color online) Example of a directed graph representing the
all microstates and transitions between them for the Q1D system with
$N=7$, $J=2$, $K=1$, and $M=2$ near $\phi^*$. Diamonds and octagons represent
microstates with only uncompressed regions and microstates with
at least one compressed region, respectively. The orange-shaded symbols are
microstates that occur in the bottleneck-crossing pathway shown in
Fig.~\ref{N7}.}
\label{complex network}
\end{figure*}

In our approach, the set of microstates and transitions between them
(for a given $\phi$ near kinetic arrest) are represented by a directed
graph.  The microstates correspond to nodes of the graph, and the
transitions between states correspond to the edges connecting the
nodes.  This graphical representation is illustrated in Figs.\
\ref{fig-8-network} and \ref{complex network} for systems with a
single and two intersections, respectively.

Transitions between two microstates occur when a particle crosses
a border between a lobe and an intersection (see Fig.\ \ref{transition
  between states}).  Since the particle can cross the border in a
positive or negative direction (cf.\ the definition in
Fig.\ \ref{system geometry}), the transitions are represented by
directed edges depicted as arrows (the arrow orientation corresponds
the positive direction of particle motion).

Our goal is to identify bottleneck microstates (microstate clusters) that
control the slow dynamics of the system at long times.  Thus, in our graphical
representation we distinguish nodes corresponding to unsqueezed
and squeezed microstates.

\subsubsection{Loops, winding number, and diffusive motion}
\label{Loops, winding number, and diffusive motion}

Since the Q1D models considered in our study involve single-file
particle arrangements, diffusive relaxation involves dynamics in
which, repeatedly, all particles are shifted either forward or
backward by approximately one position in the sequence.  Bottleneck
microstates and microstate clusters that need to be traversed to achieve a
shifted particle configuration control the slow dynamics at long times
near $\phi^*$.  Our goal is to identify and characterize such microstates.

To determine a sequence of particle displacements required to generate
cooperative translation of a single file of particles along the
channel, we focus on a set of cyclic paths on the microstate graph.
Such cyclic paths (closed loops) correspond to dynamic processes where the
systems undergoes a sequence of transitions after which it returns to
a microstate with the same state occupancy variable $\occupancyVar$ as
the initial one.  It is sufficient to consider minimal paths where
each state is visited only once, because all other paths can be
represented as a superposition of the minimal paths.

Closed loops that correspond to particle displacements that contribute
to diffusive relaxation of the system have particle labels in the
final microstate that are shifted by one in the positive or negative
direction compared to the initial microstate.  Whether particles
undergo a collective displacement (that contributes to long-time
diffusive motion) when traversing a given closed loop can be
determined by calculating the winding number,
\begin{equation}
\label{winding}
W=\sum_{i=1}^{n_p} w_i,
\end{equation}
where $i=1, 2, \ldots, n_p$ represents the sequence of transitions
between neighboring microstates, $n_p$ is the total number of
transitions in path $p$, and $w_i=\pm 1$ is the weight of transition
$i$.  The weight $w_i=1$ is assigned if the transition between two
adjacent states occurs in the direction indicated by the arrow, and
the weight $w=-1$ is assigned otherwise.  Since $W$ is incremented by
$\pm2$ when a particle crosses an intersection, and each intersection
must be crossed twice on a loop (once in mode \textit{I} and once in
mode \textit{II}), the winding number is $W=\pm4J$ when the particle
label sequence is shifted by $\pm1$ position.  Our goal is to
enumerate closed loops with nonzero winding number and determine which
loop corresponds to the shortest evolution timescale
$t_{\scriptsize{\textrm{min}}}$.  The long-time diffusive
relaxation timescale is determined as
$t_D=t_{\scriptsize{\textrm{min}}}$.

\subsection{Determination of bottleneck microstates that control the system dynamics}
\label{Determination of bottleneck states that control the system dynamics}

To identify bottleneck microstates that control the system dynamics and
determine their crossing-frequency exponents, we construct the graph
representing the microstate network in a hierarchical way.  We first
divide the set of all squeezed microstates $\squeezedSet$ into an ordered
sequence $\squeezedSet_i$, $i=1,2,\ldots,i_{\maxx}$, of subsets
containing microstates with crossing-frequency exponents
$\crossingExponent_i$ that satisfy
$\crossingExponent_1<\crossingExponent_2<\ldots<\crossingExponent_{\maxx}$.
Next, we construct a sequence of partial networks
\begin{equation}
\label{partial network}
\network_j=\bigcup_{i=0}^j\squeezedSet_i
\end{equation}
that include the set of unsqueezed microstates
$\unsqueezedSet\equiv\squeezedSet_0$ and squeezed microstates of the order
$i\le j$ (as well as all their connections).  For each subnetwork
$\network_j$, $j=0,1,2,\ldots$, we enumerate all minimal closed loops and
search for a loop with nonzero winding number.  The process stops at
the level $j=j_0$ where the first such closed loop is identified.  The
timescale for long-time diffusive relaxation is then given by 
\begin{equation}
\label{scaing of relaxation time}
t_D \sim f_{j_0}^{-1}\sim(\phiKA-\phi)^{-\crossingExponent_{j_0}},
\end{equation}
where $f_{j_0}$ is the crossing frequency \eqref{frequency of state
  crossing general} with exponent $\crossingExponent=\crossingExponent_{j_0}$.

In our numerical implementation of the above procedure, the microstate networks
were created using the C++ boost graph library~\cite{cplusplus,boost}
in combination with Python graph libraries and visualized using
the \textsc{Graphviz} graph visualization software~\cite{graphviz}.  To
simplify closed loop enumeration, some microstates were discarded based on
geometrical considerations showing that they cannot participate in a
loop that controls long-time diffusive dynamics.

\begin{figure}
\includegraphics*[width=0.3\textwidth]{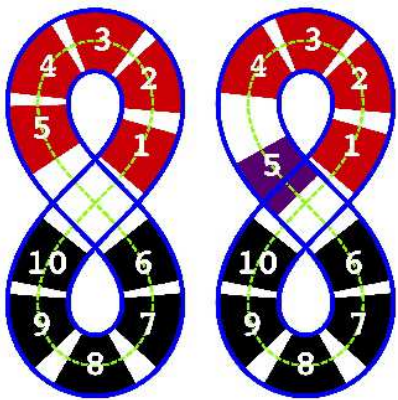}
\caption{\label{transition between states} Illustration of a
transition from one microstate to another, from $S=505$ to $534$
(i.e. in the positive direction), in a Q1D system with $N=10$ and
$J=1$.  Before the transition, particle $5$ is associated with the top
end lobe, and after the transition it is associated with the junction.
}
\end{figure}

%%%%%%%%%%%%%%%%%
\begin{figure*}
\begin{center}
\scalebox{0.6}{\includegraphics{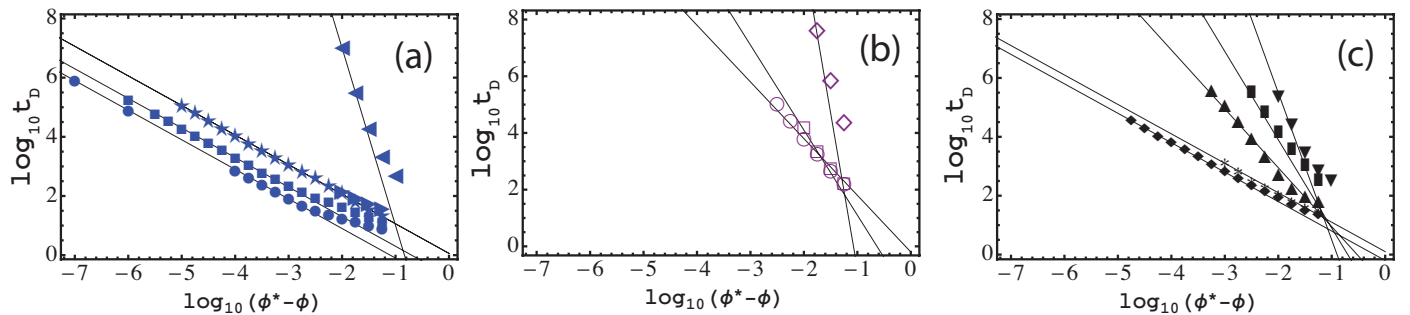}}
\end{center}
\vspace{-0.6in}
\caption{\label{Figure-16DiffusionTimes} Structural relaxation time
$t_D$ plotted versus distance from kinetic arrest $\phi^*-\phi$ for
$13$ different Q1D topologies: (a) $J=2$ and $K=1$, (A)-(E) in
Table~\ref{simulation parameters}; (b) $J=2$ and $K=2$, (G)-(I) in
Table~\ref{simulation parameters}; and (c) $J=3$ and $K=1$, (J)-(M),
and $J=4$ and $K=1$, (N).  The slopes of the solid black lines
correspond to the dominant power-law scaling exponent $\alpha$
predicted by Eqs.~\eqref{crossing exponent U} and~\eqref{crossing
exponent T}, and shown in Table~\ref{simulation parameters}.}
\vspace{0.0in}
\end{figure*}
%%%%%%%%%%%%%%%%%

\begin{table}
\newcommand{\crr}{&\quad}
\newcommand{\ON}[2]{#1}

\begin{tabular}{crrrrrcr}
\hline\hline
Case\crr $\junctions$ \crr $\Klobe$ \crr $\Nparticles$ \crr $\Mvoids$ \crr $\Nc$ \crr Type \crr $\crossingExponent$ \\
\hline
A   &      2       &     1    &        4      &     4     &    2  &  S   &       1\\\\
B   &      2       &     1    &        5      &     3     &    2  &  S   &       1\\\\
C   &      2       &     1    &        6      &     2     &    2  &  S   &       1\\
     &              &          &               &           &    1  &  T   &       1 \\\\
D   &      2       &     1    &        7      &     1     &    3  &  U   &       1\\
     &              &          &               &           &    2  &  S   &       1 \\
     &              &          &               &           &    1  &  T   &       1 \\\\
E   &      2       &     1    &        8      &     0     &    8  &  U   &       6\\\\
%F?   &      2       &     2    &       11      &     3     &    2  &  T   &       2\\\\
\ON{G}{F}   &      2       &     2    &       12      &     2     &    4  &  U   &       2\\
     &              &          &               &           &    2  &  T   &       2 \\\\
\ON{H}{G}   &      2       &     2    &       13      &     1     &    5  &  U   &       3\\\\
\ON{I}{H}   &      2       &     2    &       14      &     0     &   13  &  U   &      11\\\\
\ON{J}{I}   &      3       &     1    &        7      &     4     &    2  &  S   &       1\\\\
\ON{K}{J}   &      3       &     1    &        8      &     3     &    3  &  U   &       1\\
     &              &          &               &           &    2  &  S   &       1 \\\\
\ON{L}{K}   &      3       &     1    &        9      &     2     &    4  &  U   &       2\\\\
\ON{M}{L}   &      3       &     1    &       10      &     1     &    5  &  T   &       5\\\\
\ON{N}{M}   &      4       &     1    &       12      &     2     &    5  &  U   &       3\\
\hline\hline

\end{tabular}

\caption{\label{simulation parameters}Parameters of $13$ systems for which
  results for the structural relaxation time $t_D$ are presented in Fig.\ \ref{Figure-16DiffusionTimes}.}
\end{table}

\section{Comparison of results of bottleneck microstate analysis with Monte Carlo simulations}
\label{Bottleneck states ad long time diffusion}

To verify the conceptual framework described in Sec.\ \ref{Discrete
microstates and microstate network}, we performed microstate analyses
for a variety of Q1D systems with different topologies and carried out
extensive Monte Carlo simulations to measure the long-time diffusive
relaxation time as a function of $\phi^*-\phi$. A summary of our
results is presented in Table \ref{simulation parameters} and Fig.\
\ref{Figure-16DiffusionTimes}.  For systems denoted
C--E in Table \ref{simulation parameters} we have performed a complete
network analysis, and for the remaining cases the results are based on
geometrical investigations of particle motion to
identify bottleneck microstates.

\begin{figure}
\includegraphics*[width=0.5\textwidth]{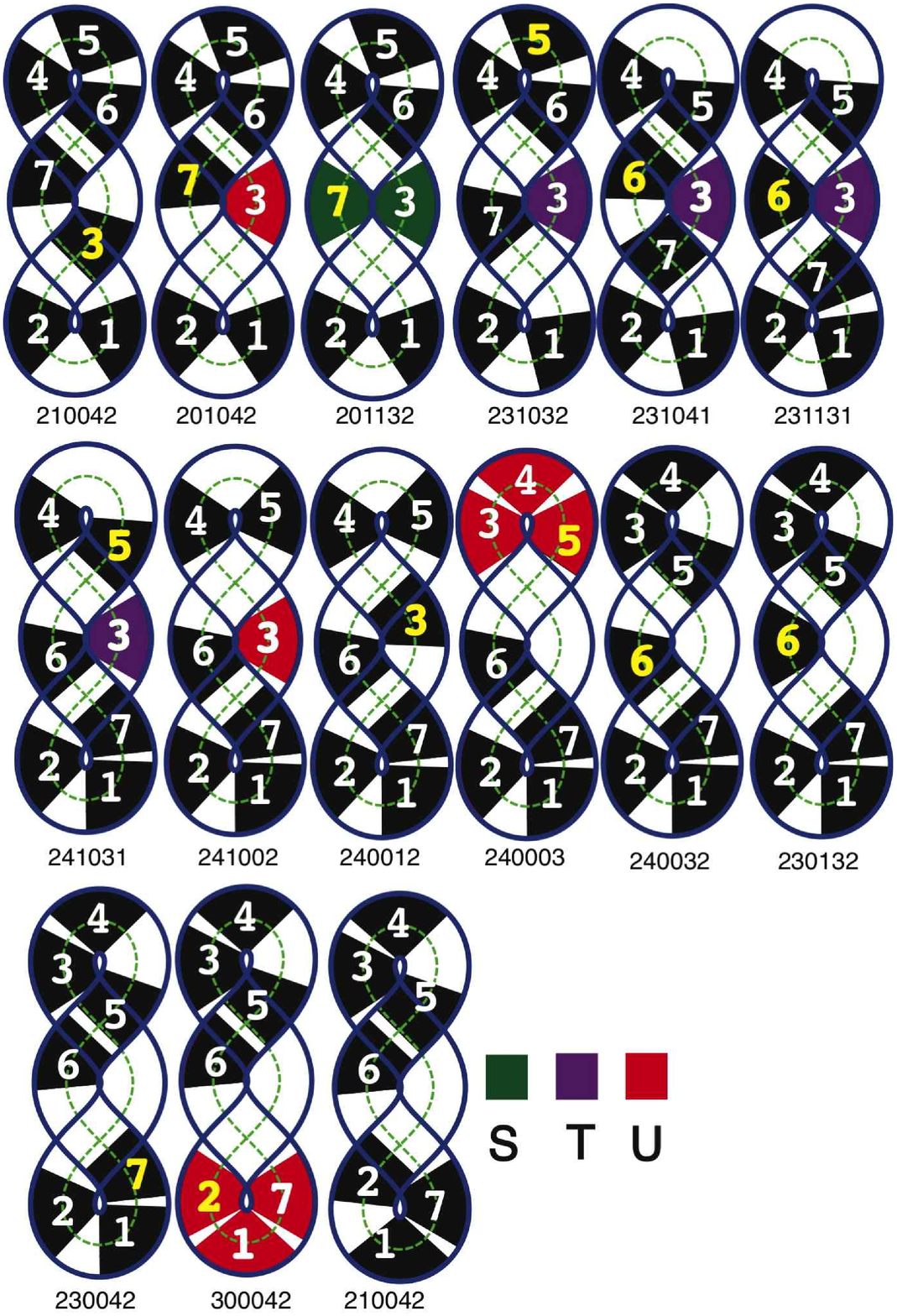}
\caption{Microstate pathway for $N=7$, $J=2$, $K=1$, and 
$M=2$ that moves from microstate $210042$ through a sequence 
of bottleneck microstates and back to microstate $210042$ with the 
particle labels shifted forward by one. The highlighted particle 
labels indicate the particle that will move in the next frame.  Green-, 
violet-, and red-shaded particles represent semi-trapped, trapped, 
and untrapped compressed regions, respectively, that control the 
crossing-frequency exponent $\alpha$.}
\label{N7}
\end{figure}

\begin{figure}
\includegraphics*[width=0.5\textwidth]{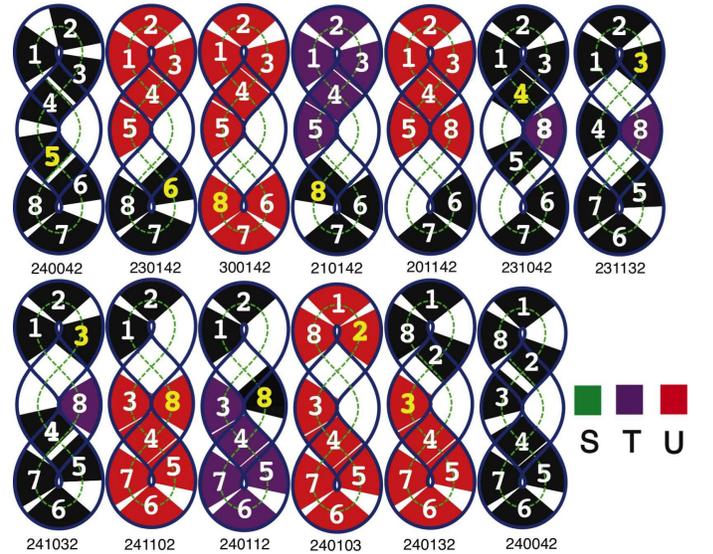}
\caption{Microstate pathway for $N=8$, $J=2$, $K=1$, and $M=1$ that
moves from microstate $240042$ through a sequence of bottleneck
microstates and back to microstate $210042$ with the particle labels
shifted forward by one. The highlighted particle labels indicate the
particle that will move in the next frame.  Violet- and red-shaded
particles represent trapped and untrapped compressed regions,
respectively, that control the crossing-frequency exponent $\alpha$.}  
\label{N8} 
\end{figure}

Table \ref{simulation parameters} shows the number of compressed
particles $\Nc$ in the bottleneck microstates that were detected for a
given system topology and indicates if a given microstate is untrapped
(U), semitrapped (S), or trapped (T).  The resulting
crossing-frequency exponent $\crossingExponent$ is also given.  Two
examples of complete pathways that enable the particle labels of
strings of particles to move forward or backward by one are depicted
in Figs.\ \ref{N7} and \ref{N8} for systems D and E, respectively.
The figures show displacements of individual particles that lead to
each microstate transition.  In addition, the compressed regions in
the controlling bottleneck microstates are marked.  In case D, the
system goes through three types of bottleneck microstates:
(\textit{a}) a microstate with a one-particle trapped CR, (\textit{b})
a microstate with a two-particle semitrapped CR, and (\textit{c}) a
microstate with a three-particle untrapped CR.  All of these
microstates yield the same crossing-frequency exponent
$\crossingExponent=1$ for this topology.  In case E, the evolution of
the system is dominated by bottleneck microstates with all eight
particles forming the compressed region.  Since the bottleneck
microstates are untrapped, we have $\crossingExponent=6$ for this
topology.

The scaling behavior of the long-time structural relaxation time
predicted by our bottleneck microstate analysis is compared in
Fig.\ \ref{Figure-16DiffusionTimes} with results of direct Monte Carlo
simulations of the system.  In all cases, we find that the
simulation results are consistent with the results of our theoretical
analyses.  (We note, however, that due to extremely long simulation
times required to determine the critical dynamics near KA, in not all
cases the simulations allow a unique determination of the critical
exponent.)

%%%%%%%%%%%%%%%%%
\begin{figure}[h]
\begin{center}
\scalebox{0.45}{\includegraphics{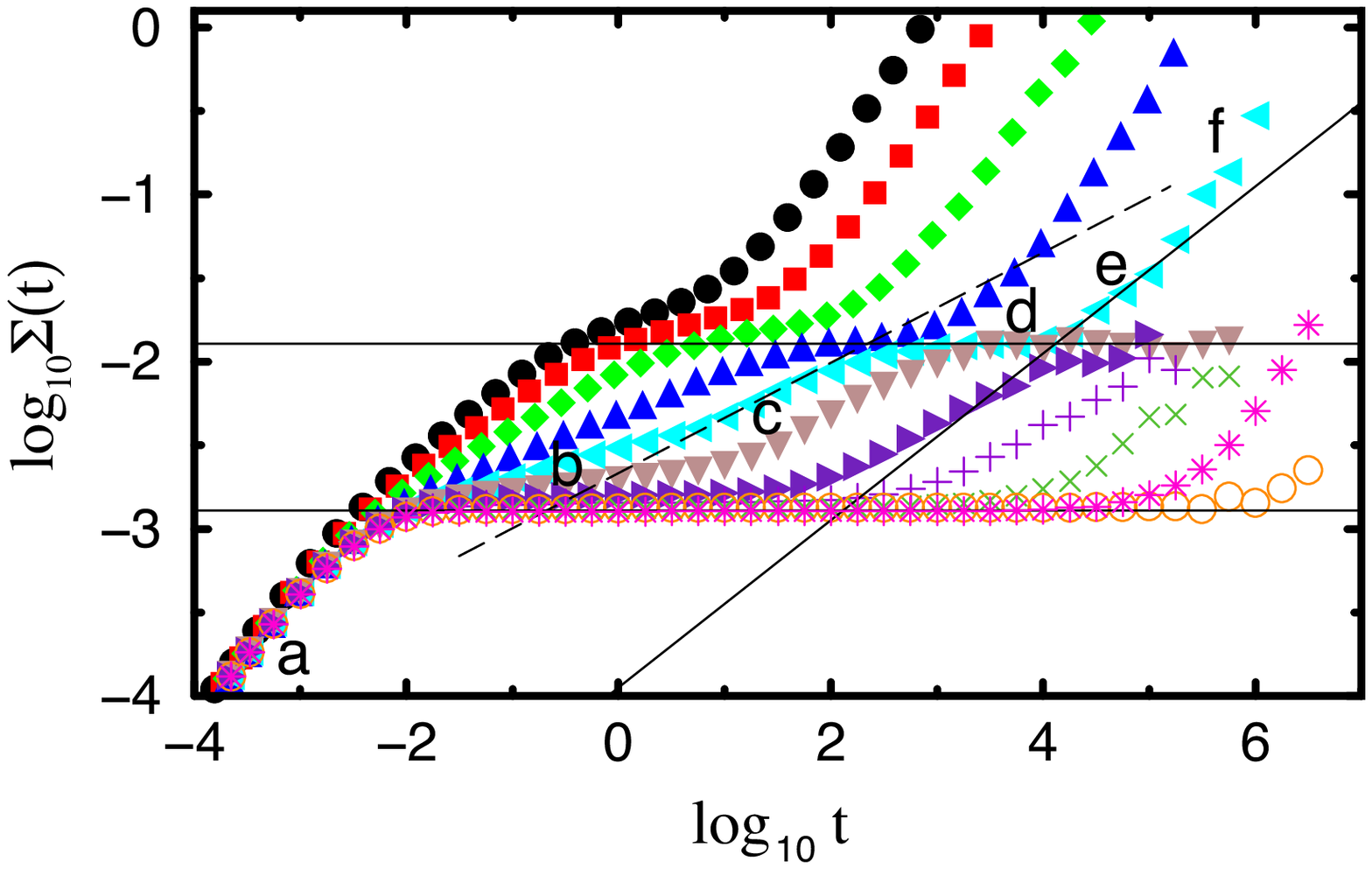}}
\end{center}
\vspace{-0.2in}
\caption{\label{J3K1N10msd} Mean-square displacement $\Sigma(t)$
versus time $t$ for the Q1D model with $N=10$, $J=3$, and $K=1$. The
packing fraction is varied from $\phi=0.499$ (filled circles) to
$0.555$ (open circles) from left to right. The packing fraction 
at kinetic arrest, $\phi^* = 0.625$, for
this topology. The labels a-f correspond to a)
short-time diffusive behavior, b) formation of a short-time plateau,
c) sub-diffusive behavior, d) formation of second intermediate-time
plateau, e) a second sub-diffusive regime, and f) long-time diffusive
behavior, respectively.  The solid horizontal lines at
$\log_{10}\Sigma(t) \approx -2.90$ and $\log_{10}\Sigma(t) \approx
-1.90$ indicate the localization lengths for first and second
plateaus, respectively. The dashed and solid lines provide the slopes,
$0.33$ and $0.5$, of the first and second sub-diffusive regimes.}
\vspace{-0.1in}
\end{figure}
%%%%%%%%%%%%%%%%%

\section{Conclusions and Future Work}
\label{conclusions}

In this manuscript, we analyzed and performed Monte Carlo simulations
of a quasi-one-dimensional model in which hard rods undergo
single-file Brownian motion inside a series of intersecting narrow
channels.  Like supercooled liquids and glasses, this Q1D model
displays slow and cooperative dynamics as the packing fraction
approaches $\phi^*$, which signals complete kinetic arrest and varies
with the system topology.  We provided a complete analysis 
of the model dynamics for several different topologies beyond the 
`figure-8' model described previously~\cite{prasanta}. 

We mapped each configuration of particles to a set of
discrete microstates that describe the number of particles in each
lobe and occupancy of the intersections.  For several system
topologies, we enumerated all microstates near $\phi^*$ and
constructed directed graphs that identify all transitions between
microstates. We find that Q1D systems must pass through a set of rare
`bottleneck' microstates to reach the long-time diffusive regime.  The
time required to reach the diffusive regime grows as a power-law, $t_D
\sim (\phi^* - \phi)^{-\alpha}$, as the packing fraction approaches
$\phi^*$ with an exponent $\alpha$ that is determined by the system
topology.  Note that since the packing fraction of kinetic arrest
$\phi^*$ and the associated structural relaxation time can be calculated
exactly, it is straightforward to fully equilibrate the
system at each packing fraction.

We have identified several intriguing features of the dynamics of Q1D
systems that require further investigation.  First, our current
studies have been limited to rather small numbers of particles $N$ and
junctions $J$.  As shown in Fig. 4 (b), for small $N$ and $J$, one is not
able to obtain single-file diffusive behavior $\Sigma(t) \sim
t^{\beta}$, with $0 < \beta < 1$ over a wide dynamical range.

To illustrate subdiffusive behavior in Q1D systems beyond the figure-8
topology, we carried out preliminary studies of the MSD for a system
with $N=10$, $J=3$, and $K=1$ in Fig.~\ref{J3K1N10msd}. Even though
the packing fraction in Fig.~\ref{J3K1N10msd} is significantly below
$\phi^*$, we observe multiple plateaus and regions of subdiffusive
behavior.  The exponents of the subdiffusive behavior in regions c and
e are approximately $0.33$ and $0.5$, respectively.  The slope of
$0.5$ indicates possible single-file diffusive
behavior~\cite{nelissen,ryabov}.  In future studies, we will investigate systems
with increasing numbers of particles at fixed topology to test the
robustness of the subdiffusive behavior, and identify rare microstates
that give rise to structural relaxation from the caged regions b and
d.  These preliminary results indicate that there is nontrivial dynamics in Q1D
systems even far from kinetic arrest.

We believe that this work will encourage new simulations and
experiments of dense colloidal and other glassy systems in narrow channels
and also in bulk to determine whether bulk systems can display
quasi-one-dimensional dynamical behavior and whether the effective Q1D
topology of the system can vary significantly with packing fraction.

\section*{Acknowledgments}
Financial support from the NSF CBET-1059745 (J.B.) and DMR-1006537 (C.O.
and P.P.) is gratefully acknowledged. This work also benefited from the facilities and staff of the Yale University Faculty of Arts and Sciences High 
Performance Computing Center and the NSF (Grant No. CNS-0821132) that 
in part funded acquisition of the computational facilities. 

\bibliographystyle{ieeetr}
\bibliography{Q1D}

\begin{thebibliography}{10}
\bibitem{rcp}
Random close packing for these systems are $\phi_{\rm rcp} = 0.62$ and $0.63$
  for $6\%$ and $10\%$ polydispersities, respectively.
\end{thebibliography}
\end{document}